\documentclass[11pt, fceqn]{article}
 \usepackage{amsfonts}
 \usepackage{flafter}
 \usepackage{amssymb,graphics,graphicx,subfigure,caption2,rotating}
 \usepackage{amsmath, latexsym}
\usepackage[amsmath,thmmarks]{ntheorem}
 \usepackage[top=3.5cm]{geometry}
\begin{document}
\date{}
\title{Improved separability criteria via some classes of measurements}
\author{Shu-Qian Shen$^{1}$\thanks{Email: sqshen@upc.edu.cn}, Ming Li$^1$\thanks{Email: liming3737@163.com}, Xianqing Li-Jost$^{2}$\thanks{Email: xli-jost@mis.mpg.de}, Shao-Ming Fei$^{2,3}$\thanks{Email: feishm@cnu.edu.cn}\\
{\small{\it $^1$College of Science, China University of Petroleum, 266580 Qingdao, P.R. China}}\\
{\small{\it $^2$Max-Planck-Institute for Mathematics in the Sciences, 04103 Leipzig, Germany}}\\
{\small{\it $^3$School of Mathematical Sciences, Capital Normal University, 100048 Beijing, P.R. China}}}\maketitle
\begin{abstract}

The entanglement detection via local measurements can be experimentally implemented. Based on mutually unbiased measurements and general symmetric informationally complete positive-operator-valued measures, we present separability criteria for bipartite quantum states, which, by theoretical analysis, are stronger than the related existing criteria via these measurements. Two detailed examples are supplemented to show the efficiency of the presented separability criteria.
\end{abstract}

\section{Introduction}
Entanglement is one of the main differences between quantum mechanics and classical physics. As an essence resource, it has been widely used to implement some quantum tasks from quantum cryptography to  quantum teleportation; see, e.g., \cite{Nielsen2010}. Thus, the detection or determination of entanglement of any quantum state becomes extremely important and necessary. Although many efforts have been devoted to the study of this problem, it is still open except for the case of $m\times n$ quantum states with $mn\le 6$ \cite{ppt,ppt1,Horodecki1997}. Nevertheless, in the last years, a variety of sufficient conditions for entanglement have been proposed; see \cite{survey} for a survey.

Among them, the separability criteria based on quantum measurements are attractive in the last years, since they are easily experimentally implemented.
In  \cite{Spengler2012}, the mutually unbiased bases (MUBs) \cite{Schwinger1960} were connected with the entanglement detection for bipartite, multipartite and continuous-variable quantum systems. The detection ability of the presented criteria partly depends on the maximum number of MUBs.   For $\mathbb{C}^d$, if the  dimension $d$ is an integer power of a prime number, then extremal sets containing $d+1$ MUBs are known. However, for arbitrary $d$ the maximum number of MUBs is still not known \cite{Durt2010}.

  In \cite{Kalev2014-1}, MUBs were generalized to mutually unbiased measurements (MUMs). Unlike MUBs, a complete set of MUMs, i.e., $d+1$ MUMs, can always be constructed with certain efficiency parameters for any $d$-dimensional space. Based on MUMs, Chen, Ma and Fei \cite{Chen2014-1} obtained a new separability criterion including the corresponding criterion in \cite{Spengler2012} as a special case. In \cite{Shen2015}, by applying MUMs to $\rho-\rho^A\otimes \rho^B$, where $\rho^A$ and $\rho^B$ are the reduced density matrices of $\rho$, the authors obtained a new separability criterion, which is more powerful than the corresponding ones in \cite{Spengler2012} and \cite{Chen2014-1}. Later, these criteria were further generalized to the multipartite case \cite{Liu2015-1,Liu2015-2,Liu2017,Lu2017}. It should be mentioned that the criteria in \cite{Liu2015-1,Liu2017,Lu2017} can be used to the quantum systems with subsystems having different dimensions. Recently, Rastegin \cite{Rastegin2016} derived separability criteria in terms of local fine-grained uncertainty relations under MUBs and MUMs.

    Another kind of quantum measurements, known as symmetric informationally complete positive-operator-valued measures (SIC-POVMs)  \cite{Wootters2004}, has a close connection with MUBs. In \cite{Kalev2014-2}, a family of  general SIC-POVMs (GSIC-POVMs) has been constructed. Unlike SCI-POVMs, the measurement operators of GSIC-POVMs are not necessarily rank $1$. By analogous arguments to MUMs, the separability criteria based on GSIC-POVMs were also investigated in \cite{Shen2015,Liu2015-2,Xi2016,Chen2015,Lu2017}.

  The aim of this paper is to achieve some new separability criteria via MUMs and GSIC-POVMs. They are strictly stronger than the corresponding ones in \cite{Spengler2012,Chen2014-1,Shen2015,Liu2015-1,Xi2016,Chen2015,Lu2017}.

  The remainder of the paper is organized as follows. In Section 2, some preliminaries about MUMs and GSIC-POVMs are briefly reviewed. In Section 3, the separability criteria based on MUMs are obtained for bipartite quantum states. In Section 4, by similar methods used in Section 3, the separability criteria via GSIC-POVMs for bipartite quantum states are derived. Some conclusions and future work are drawn in Section 5.

\section{Some classes of measurements}

The concept of mutually unbiased bases (MUBs) was first introduced by Schwinger \cite{Schwinger1960}. The two orthonormal bases $\{|a_i\rangle\}_{i=1}^d$ and $\{|b_i\rangle\}_{i=1}^d$ of $\mathbb{C}^d$ are said to be mutually unbiased, if and only if
 \begin{equation*}
 |\langle a_i|b_j\rangle|^2=\frac{1}{d},i,j=1,\cdots,d.
 \end{equation*}
A set of orthonormal bases are MUBs if each two of them are mutually unbiased. The MUBs have key applications in quantum information processing such as quantum state tomography, mean king¡¯s problems, quantum cryptography; see, e.g., \cite{Durt2010}. When $d$ is an integer power of a prime number, a complete set of MUBs, i.e., $d+1$ MUBs, can be constructed \cite{Wootters1989}. However, whether there exists a complete set of MUBs for any $d$ is still unknown.

Mutually unbiased measurements (MUMs) \cite{Kalev2014-1} can be seen as a generalization of MUBs under weaker requirements. Two measurements on $\mathbb{C}^d$
\[
\mathcal{P}^{(b)}=\left\{P_n^{(b)}|P_n^{(b)}\ge 0,\sum\limits_{n=1}^d P_n^{(b)}=I_d\right\}, b=1,2
\]
are said to be mutually unbiased if and only if
\begin{description}
  \item[(i)] Tr($P_n^{(b)})=1$;
  \item[(ii)] Tr$(P_n^{(b)}P_{ n'}^{(b')})=\delta_{nn'}\delta_{bb'} \kappa +(1-\delta_{nn'})\delta_{bb'}\frac{1-\kappa}{d-1}+(1-\delta_{bb'})\frac{1}{d}$,
          \end{description}
where the efficiency parameter $\kappa$ satisfies $\frac{1}{d}<\kappa\le 1$.\emph{ A complete set of MUMs, i.e., $d+1$ MUMs, can be constructed in any $d$-dimensional space; see} \cite{Kalev2014-1} \emph{for a detail.} If $\kappa$ can be chosen to be $\kappa=1$, a complete set of MUMs reduces to a complete set of MUBs.

 A set of $d^2$ operators
 \[
\left\{M_i|M_i\ge 0,\sum\limits_{i=1}^{d^2}M_i=I_d\right\}
\]
is said to be a general symmetric informationally complete positive-operator-valued measure (GSIC-POVM) \cite{Kalev2014-2} if and only if
\begin{description}
  \item[(i)] Tr$(M^2_i)=\alpha, i=1,\cdots,d^2$;
  \item[(ii)] Tr$(M_i M_{ j})=\frac{1-d\alpha}{d(d^2-1)},1\le i\neq j\le d^2$,
\end{description}
where the parameter $\alpha$ satisfies $\frac{1}{d^3}<\alpha\le \frac{1}{d^2}$. The complete set of GSIC-POVMs can be constructed explicitly in all finite dimensions for some choices of $\alpha$ \cite{Kalev2014-2}. If $\alpha$ can be chosen to be $\alpha=\frac{1}{d^2}$, then the complete set of GSIC-POVMs becomes the complete set of SIC-POVMs \cite{Renes2004}.

\section{Entanglement detection via MUMs}
In this section, based on MUMs, we give separability criteria for bipartite states. In the following, for any matrix $A$, we denote by $\text{Tr}(A)$, $A^T$, $||A ||_{\text{tr}}$ and $||A||_2$ the trace, the transpose, the trace norm (i.e., the sum
of singular values) and the spectral norm (i.e., the maximum singular
value) of $A$, respectively.

Let ${\{\mathcal{P}^{(b)}\}}_{b=1}^{m_1}$ and ${\{\mathcal{Q}^{(b)}\}}_{b=1}^{m_2}$ be two sets of MUMs on $\mathbb{C}^{d_1}$ and $\mathbb{C}^{d_2}$, respectively.  Then we define $X=\{X_i\}_{i=1}^{d_1m_1}$ and $Y=\{Y_j\}_{j=1}^{d_2m_2}$ with
\begin{align}
\label{X1}
X_i=P_{n_1}^{(b_1)},i=(b_1-1)d_1+n_1,b_1=1,\cdots,m_1,n_1=1,\cdots,d_1,\\
\label{X2} Y_j=Q_{n_2}^{(b_2)},j=(b_2-1)d_2+n_2,b_2=1,\cdots,m_2,n_2=1,\cdots,d_2.
\end{align}
For any state $\rho$ in $ \mathbb{C}^{d_1}\otimes\mathbb{C}^{d_2}$, we further define
\begin{equation*}
\label{measuredmatrix}
\mathcal{M}^{(X,Y)}(\rho)=(w_{ij})\in\mathbb{C}^{d_1m_1\times d_2m_2}
\end{equation*}
with
\[
w_{ij}=\text{Tr}\left(X_i\otimes Y_j(\rho-\rho^A\otimes \rho^B)\right),
\]
where $\rho^A$ and $\rho^B$ are the reduced density matrices acting on the first and second subsystems, respectively. The following theorem gives the new separability criterion based on $\mathcal{M}^{(X,Y)}(\rho)$.
\\
\\
\textbf{Theorem 3.1.} \emph{Let ${\{\mathcal{P}^{(b)}\}}_{b=1}^{m_1}$ and ${\{\mathcal{Q}^{(b)}\}}_{b=1}^{m_2}$ be two sets of MUMs  on $\mathbb{C}^{d_1}$ and $ \mathbb{C}^{d_2}$ with efficiency parameters $\kappa_1$ and $\kappa_2$, respectively, and let $X=\{X_i\}_{i=1}^{d_1m_1}$ and $Y=\{Y_j\}_{j=1}^{d_2m_2}$ be defined as in \emph{(\ref{X1})} and \emph{(\ref{X2})}. If the quantum state $\rho$ in $\mathbb{C}^{d_1}\otimes \mathbb{C}^{d_2}$ is separable, then}
\begin{equation}
\label{th331}\left\|\mathcal{M}^{(X,Y)}(\rho)\right\|_{\text{tr}}\le \sqrt{\frac{m_1-1}{d_1}+\kappa_1-\sum\limits_{i=1}^{d_1m_1}\left(\text{Tr}(X_i\rho^A)\right)^2}\sqrt{\frac{m_2-1}{d_2}+\kappa_2-\sum\limits_{i=1}^{d_2m_2}\left(\text{Tr}(Y_i\rho^B)\right)^2}.
\end{equation}

See the ``Appendix" for the proof. \\

In general, the criterion given by Theorem 3.1 is more efficient when $m_i$ gets larger from $1$ to $d_i+1$, $i=1,2$; see, e.g., \cite{Spengler2012}. Hence, consider the cases $d_1=d_2=d$, $m_1=m_2=d+1$, and $\kappa_1=\kappa_2=\kappa$. (\ref{th331}) in Theorem 3.1 reduces to
\begin{equation}
\label{con}||\mathcal{M}^{(X,Y)}(\rho)||_{\text{\text{tr}}}\le \sqrt{1+\kappa-\sum\limits_{i=1}^{d(d+1)}\left(\text{Tr}(X_i\rho^A)\right)^2}\sqrt{1+\kappa-\sum\limits_{i=1}^{d(d+1)}\left(\text{Tr}(Y_i\rho^B)\right)^2}.
\end{equation}
In this case, the criterion \cite[Theorem 2]{Shen2015} states that any separable state $\rho$ in $ \mathbb{C}^{d}\otimes \mathbb{C}^{d}$ satisfies
\begin{equation}
\label{Shen2015j}\sum\limits_{i=1}^{d(d+1)}\left|\text{Tr}\left(X_i\otimes Y_i(\rho-\rho^A\otimes\rho^B)\right)\right|
\le \sqrt{1+\kappa-\sum\limits_{i=1}^{d(d+1)}\left(\text{Tr}(X_i\rho^A)\right)^2}\sqrt{1+\kappa-\sum\limits_{i=1}^{d(d+1)}\left(\text{Tr}(Y_i\rho^B)\right)^2},
\end{equation}
which is weaker than (\ref{con}) by the inequality \cite{Vicente2007}
\begin{equation*}
\label{vi}\sum\limits_{i=1}^n\left|g_{ii}\right|\le ||G||_{\text{tr}}\text{ for any matrix } G=(g_{ij})\in \mathbb{C}^{n\times n}.
\end{equation*}
Since the criteria given in \cite{Spengler2012,Chen2014-1} are weaker than (\ref{Shen2015j}), the criterion (\ref{con}) from Theorem 3.1 is the strongest one among these criteria.

Recently, Liu et al. \cite{Liu2015-1} presented separable criteria for quantum states with different dimensions of subsystems. Let $d=\min\{d_1,d_2\}.$ Consider $m_1=m_2=m$. We define two subsets
\[
\{p_1,p_2,\cdots,p_{md}\}\subseteq \{1,2,\cdots,md_1\},~~ \{q_1,q_2,\cdots,q_{md}\}\subseteq \{1,2,\cdots,md_2\}\]
 where
\begin{align*}
id_1+1\le p_{id+l}\le (i+1)d_1,~~& jd_2+1\le q_{jd+l}\le (j+1)d_2
\end{align*}
for $l=1,\cdots,d$ and $ i,j=0,1,\cdots,m-1$.
Set $G=(g_{ij})\in \mathbb{C}^{md\times md}$  with $g_{ij}=\text{Tr}(X_{p_i}\otimes Y_{q_j}\rho)$. The criterion \cite[Theorem 3]{Liu2015-1} shows that any separable state $\rho$ in $\mathbb{C}^{d_1}\otimes\mathbb{C}^{d_2}$ satisfies
\begin{equation}
\label{cri-Liu}
\text{Tr}(G)=\sum\limits_{i=1}^{md}\text{Tr}(X_{p_i}\otimes Y_{q_i}\rho)\le  \sqrt{\frac{m-1}{d_1}+\kappa_1}\sqrt{\frac{m-1}{d_2}+\kappa_2}.
\end{equation}
 If $W$ is a sub-matrix of the matrix $V$, then from \cite{Stewart1990} we have
  $||W||_{\text{tr}}\le ||V||_{\text{tr}}$. By this conclusion and similar comparison between \cite{Chen2014-1} and \cite{Shen2015}, it can be found that Theorem 3.1, with $m_1=m_2=m$, is stronger than the criterion (\ref{cri-Liu}).

 We now compare Theorem 3.1 with the existing related criteria by examples.
\\
\\
\textbf{Example 3.1.} Consider the following $3\times 3$ bound entangled state \cite{Bennett1999}:
\begin{equation*}
\rho=\frac{1}{4}\left(I_9-\sum\limits_{i=0}^4 |\eta_i\rangle\langle\eta_i|\right),
\end{equation*}
where
\begin{align}\nonumber
|\eta_0\rangle&=\frac{1}{\sqrt{2}}|0\rangle(|0\rangle-|1\rangle),~|\eta_1\rangle=\frac{1}{\sqrt{2}}(|0\rangle-|1\rangle)|2\rangle,~|\eta_2\rangle
=\frac{1}{\sqrt{2}}|2\rangle(|1\rangle-|2\rangle),\nonumber\\ \nonumber
|\eta_3\rangle&=\frac{1}{\sqrt{2}}
(|1\rangle-|2\rangle)|0\rangle,~ |\eta_4\rangle=\frac{1}{3}(|0\rangle+|1\rangle+|2\rangle)(|0\rangle+|1\rangle+|2\rangle).
\end{align}
Consider the mixture of $\rho$ with white noise:
\begin{equation*}
\rho_p=\frac{1-p}{9}I_9+p\rho,\;\;0\le p\le 1.
 \end{equation*}
The complete set of MUMs is always constructed by using generalized Gell-Mann operators; see \cite{Kalev2014-1} for a detail. Numerical computations show that the existing criteria \cite{Spengler2012,Chen2014-1,Shen2015} based on MUMs cannot detect any entanglement of $\rho_p$. The criterion (\ref{con}) can detect entanglement of $\rho_p$ for $0.8822\le p\le 1$. Thus, (\ref{con}) is the best one among these criteria.
\\
\\
\textbf{Example 3.2.} The following $d\times d$ Werner states are due to \cite{Werner1989}:
\[
\eta_r=\frac{1}{d^3-d} \left((d-r)I_{d^2}+(dr-1)\sigma\right),
\]
where $-1\le r\le 1$, $\sigma=\sum\nolimits_{i,j=0}^{d-1}|ij\rangle\langle ji|$ is the ``flip" or ``swap" operator. For $d=3$, the separability criteria in \cite{Chen2015} based on MUMs cannot detect any entanglement in $\eta_r$. But, the separability criteria \cite[Theorem 2]{Shen2015} and (\ref{con}) can respectively detect entanglement of $\eta_r$ for $-1\le r\le-0.3340 $. Thus, for Werner states, the criterion (\ref{con}) is as efficient as \cite[Theorem 2]{Shen2015}. Nevertheless, they both outperform \cite[Theorem 2]{Shen2015}.

\section{Separability criteria via GSIC-POVMs}
Similar to Theorem 3.1, the separability criterion based on GSIC-POVMs can be derived for bipartite states.
\\
\\
\textbf{Theorem 4.1.} \emph{Let $P=\{P_b\}_{b=1}^{d_1^2}$ and $Q=\{Q_b\}_{b=1}^{d_2^2}$ be two sets of GSIC-POVMs on $\mathbb{C}^{d_1}$ and $\mathbb{C}^{d_2}$ with parameters $\alpha_1$ and $\alpha_2$, respectively. If the quantum state $\rho$ in $\mathbb{C}^{d_1}\otimes \mathbb{C}^{d_2}$ is separable, then}
\begin{equation*}
||\mathcal{M}^{(P,Q)}(\rho)||_{\text{tr}}\le \sqrt{\frac{\alpha_1 d_1^2+1}{d_1(d_1+1)}-\sum\limits_{i=1}^{d_1^2}(\text{Tr}(P_i\rho^A))^2}\sqrt{\frac{\alpha_2 d_2^2+1}{d_2(d_2+1)}-\sum\limits_{i=1}^{d_2^2}(\text{Tr}(Q_i\rho^B))^2}.
\end{equation*}

By analogous analysis to Theorem 3.1, one can show that Theorem 4.1 is more efficient than the corresponding criteria in  \cite[Theorem 1]{Chen2015}, \cite[Theorem 3]{Shen2015} and \cite[Theorem 2]{Xi2016}.

We now, by states in Examples 3.1 and 3.2, compare the efficiency of Theorem 4.1 and the related criteria based on GSIC-POVMs. For $\rho_p$, the criteria in \cite{Chen2014-1, Shen2015} cannot detect any entanglement for any $p$, while Theorem 4.1 can give an entanglement condition $0.8822\le p\le 1$. For $\eta_r$, using Theorem 4.1 and \cite[Theorem 3]{Shen2015}, we can obtain the same entanglement condition $-1\le r\le -0.3340$. But the criterion given in \cite{Chen2015} does not show any entanglement in $\eta_r$. In a word, Theorem 4.1 is more efficient than the corresponding criteria in \cite{Shen2015,Chen2015}.

\section{Conclusions}
Based on MUMs and GSIC-POVMs, we have derived some new separability criteria, which by theoretical analysis and numerical examples, are stronger than the corresponding existing criteria. In the future, how to investigate genuine multipartite entanglement by using MUMs and GSIC-POVMs is an interesting problem.

\section*{\bf Acknowledgments}
The authors greatly indebted to the referee and the editor for their invaluable comments and suggestions. This work is supported by the Natural Science Foundation of Shandong Province (ZR2016AM23, ZR2016AQ06), the Fundamental Research Funds for the Central Universities (18CX02035A) and the NSF of China (11675113,11775306).
\\
\begin{appendix}
\[
\textbf{Appendix: Proof of Theorem 3.1}
\]

 Since $\rho$ is separable, it can be decomposed into
\begin{equation}
\label{thm311}\rho=\sum\limits_{k=1}^r p_k\rho_k^A\otimes \rho_k^B,
\end{equation}
where $p_k>0, \sum\nolimits_kp_k=1$, $\rho_k^A$ and $\rho_k^B$ are pure states on the first and second subsystems, respectively. From the equality \cite{Zhangchenjie2008}
\[
\rho-\rho^A\otimes \rho^B=\frac{1}{2}\sum\limits_{s,t=1}^{r}p_sp_t(\rho_s^A-\rho_t^A)\otimes(\rho_s^B-\rho_t^B),
\]
we can derive
\begin{align}
\label{thm312}&\mathcal{M}^{(X,Y)}(\rho)=\frac{1}{2}\sum\limits_{s,t=1}^{r}p_sp_t\mathcal{M}\left((\rho_s^A-\rho_t^A)\otimes(\rho_s^B-\rho_t^B)\right),
\end{align}
where
\begin{align}\nonumber
&\mathcal{M}\left((\rho_s^A-\rho_t^A)\otimes(\rho_s^B-\rho_t^B)\right)=\left(\text{Tr}\left((X_i\otimes Y_j)(\rho_s^A-\rho_t^A)\otimes(\rho_s^B-\rho_t^B)\right)\right)_{d_1m_1\times d_2m_2}\\\nonumber
&=\left(\text{Tr}(X_i(\rho_s^A-\rho_t^A))\text{Tr}(Y_j(\rho_s^B-\rho_t^B))\right)_{d_1m_1\times d_2m_2}\\\nonumber
&=\left( {\begin{array}{*{20}{c}}
   {\text{Tr}({X_1}(\rho_s^A-\rho_t^A))}  \\
    \vdots   \\
   {\text{Tr}({X_{{d_1}{m_1}}}(\rho_s^A-\rho_t^A))}
   \end{array}} \right)
    \left( {\begin{array}{*{20}{c}}
   {\text{Tr}({Y_1}(\rho_s^B-\rho_t^B))} &  \cdots  & {\text{Tr}({Y_{{d_2}{m_2}}}(\rho_s^B-\rho_t^B))}\nonumber
\end{array}} \right)\\
\label{thm313}&:=\beta_{s,t}\eta_{s,t}^T.
\end{align}
From (\ref{thm312}) and (\ref{thm313}), it is easy to deduce
\begin{align*}
||\mathcal{M}^{(X,Y)}(\rho)||_{\text{tr}}&\le \frac{1}{2}\sum\limits_{s,t=1}^{r}p_sp_t||\mathcal{M}\left((\rho_s^A-\rho_t^A)\otimes(\rho_s^B-\rho_t^B)\right)||_{\text{tr}}
\\\nonumber
&=\frac{1}{2}\sum\limits_{s,t=1}^{r}p_sp_t||\beta_{s,t}\eta_{s,t}^T||_{\text{tr}}=\frac{1}{2}\sum\limits_{s,t=1}^{r}p_sp_t||\beta_{s,t}||_2||\eta_{s,t}||_{2}\\\nonumber
&\le\frac{1}{2}\sqrt{\sum\limits_{s,t=1}^{r}p_sp_t||\beta_{s,t}||_2^2}\sqrt{\sum\limits_{s,t=1}^{r}p_sp_t||\eta_{s,t}||_2^2}
\\\nonumber
&\le \sqrt{\frac{m_1-1}{d_1}+\kappa_1-\sum\limits_{i=1}^{d_1m_1}\left(\text{Tr}(X_i\rho^A)\right)^2}\sqrt{\frac{m_2-1}{d_2}+\kappa_2-\sum\limits_{i=1}^{d_2m_2}\left(\text{Tr}(Y_i\rho^B)\right)^2},\nonumber
\end{align*}
where, in the second equality, the second inequality and the third inequality, we have used, respectively, the equality
\begin{equation*}
\label{ab} |||a\rangle\langle b|||_{\text{tr}}=|||a\rangle||_2|||b\rangle||_2 \;\;\text{ for any vectors } |a\rangle \text{ and } |b\rangle,
\end{equation*}
the well-known Cauchy-Schwarz inequality, and the inequalities \cite{Rastegin2015}
\begin{align}
\label{iequalityMUM} \sum\limits_{i=1}^{d_1m_1}\text{Tr}(X_i\rho_1)^2&\le \frac{m_1-1}{d_1}+\kappa_1,~~
\sum\limits_{i=1}^{d_2m_2}\text{Tr}(Y_i\rho_2)^2\le \frac{m_2-1}{d_2}+\kappa_2
\end{align}
for any pure states $\rho_1$ and $\rho_2$ in $\mathbb{C}^{d_1}$ and $\mathbb{C}^{d_2}$, respectively.
\end{appendix}


{\small \begin{thebibliography}{99}
\bibitem{Nielsen2010} M.A. Nielsen and I.L. Chuang, \emph{Quantum computation and quantum information}, Cambridge university press, 2010.

\bibitem{ppt} A. Peres, Phys. Rev. Lett.  \textbf{77}, 1413 (1996).
 \bibitem{ppt1} M. Horodecki, P. Horodecki, and R. Horodecki,  Phys. Lett. A \textbf{223}, 1 (1996).
\bibitem{Horodecki1997} P. Horodecki, Phys. Lett. A \textbf{232}, 333 (1997).

\bibitem{survey} R. Horodecki, P. Horodecki, M. Horodecki, and K. Horodecki, Rev. Mod. Phys. \textbf{81}, 865 (2009); O. G\"{u}hne and G. T\'{o}th, Phys.  Rep. \textbf{474}, 1 (2009).
\bibitem{Spengler2012} C. Spengler, M. Huber, S. Brierley, T. Adaktylos, and B.C. Hiesmayr, Phys. Rev. A \textbf{86}, 022311 (2012).
\bibitem{Schwinger1960} J. Schwinger, Proc. Natl. Acad. Sci. USA. \textbf{46}, 570 (1960).
   \bibitem{Durt2010} T. Durt, B.G. Englert, I. Bengtsson, and K. \.{Z}yczkowski, Int. J. Quantum Inf. \textbf{8}, 535 (2010).
\bibitem{Kalev2014-1}  A. Kalev and G. Gour,  New J. Phys. \textbf{16}, 053038 (2014).
    \bibitem{Chen2014-1} B. Chen, T. Ma, and S.M. Fei, Phys. Rev. A \textbf{89},  064302 (2014).
\bibitem{Shen2015} S.Q. Shen, M. Li, and X.F. Duan, Phys. Rev. A \textbf{91}, 012326 (2015).
\bibitem{Liu2015-1} L. Liu, T. Gao, and F.L. Yan, Sci. Rep. \textbf{5}, 13138 (2015).
\bibitem{Liu2015-2} L. Liu, T. Gao, and F.L. Yan, arXiv:1512.02853 [quant-ph] (2015).
\bibitem{Liu2017} L. Liu, T. Gao, and F.L. Yan, Sci. China-Phys. Mech. Astron. \textbf{60}, 100311 (2017).
\bibitem{Lu2017} Y.Y. Lu, S.Q. Shen, T.R. Xu, and J. Yu, Int. J. Theor. Phys. \textbf{57}, 208-218 (2017).
\bibitem{Rastegin2016} A.E. Rastegin, Quantum Inf. Process. \textbf{15} 2621 (2016).
    \bibitem{Wootters2004} W.K. Wootters, arXiv:0406032v2 [quant-ph] (2004).
        \bibitem{Kalev2014-2} A. Kalev and G. Gour, J. Phys. A: Math. Theor. \textbf{47}, 335302 (2014).
\bibitem{Xi2016} Y. Xi, Z.J. Zheng, and C.J. Zhu, Quantum Inf. Process. \textbf{15}, 5119 (2016).
\bibitem{Chen2015} B. Chen, T. Li, and S.M. Fei, Quantum Inf. Process. \textbf{14}, 2281 (2015).

        \bibitem{Wootters1989} W.K. Wootters and B.D. Fields, Ann. Phys. \textbf{191}, 363 (1989).
                \bibitem{Renes2004} J.M. Renes, R. Blume-Kohout, A.J. Scott, and C.M. Caves, J. Math. Phys. \textbf{45}, 2171 (2004).
                    \bibitem{Zhangchenjie2008} C.J. Zhang, Y.S. Zhang, S. Zhang, and G.C. Guo, Phys. Rev. A \textbf{77}, 060301(R) (2008).
                               \bibitem{Rastegin2015} A.E. Rastegin, Open Syst. Inf. Dyn. \textbf{22}, 1550005 (2015).
\bibitem{Vicente2007} J.I. de Vicente, Quantum Inf. Comput. \textbf{7}, 624 (2007).
 \bibitem{Stewart1990} G.W. Stewart and J.G. Sun, \emph{Matrix perturbation theory}, Academic Press, London, 1990.
        \bibitem{Bennett1999} C.H. Bennett, D.P. DiVincenzo, T. Mor, P.W. Shor, J.A. Smolin, and B.M. Terhal, Phys. Rev. Lett. \textbf{82}, 5385 (1999).
\bibitem{Werner1989} R.F. Werner, Phys. Rev. A \textbf{40}, 4277 (1989).















%


%















\end{thebibliography}}
\end{document}